\documentclass{aa}  
\usepackage{subcaption,graphicx}
\setcitestyle{numbers}

\usepackage{graphicx}
\usepackage{txfonts}
\begin{document}

   \title{COVID-19 Pneumonia and Influenza Pneumonia Detection Using Convolutional Neural Networks}


   \author{Julianna Antonchuk \inst{1}, Benjamin Prescott \inst{1}, Philip Melanchthon \inst{1}, Robin Singh \inst{1}
          }

   \institute{\textbf{Northwestern University}\\
              juliannaantonchuk2021@u.northwestern.edu,\\ benjaminprescott2022@u.northwestern.edu,\\
              solomonmelanchthon2021@u.northwestern.edu,\\ 
              robinsingh2022@u.northwestern.edu}

 
  \abstract
   {In the research, we developed a computer vision solution to support diagnostic radiology in differentiating
between COVID-19 pneumonia, influenza virus pneumonia, and normal biomarkers. The chest radiograph appearance of COVID-19 pneumonia is thought to be
nonspecific, having presented a challenge to identify an optimal architecture of a
convolutional neural network (CNN) that would classify with a high sensitivity
among the pulmonary inflammation features of COVID-19 and non-COVID-19
types of pneumonia. Rahman (2021) states that COVID-19 radiography images
observe unavailability and quality issues impacting the diagnostic process and affecting the accuracy of the deep learning detection models. A significant scarcity
of COVID-19 radiography images introduced an imbalance in data motivating us
to use over-sampling techniques. In the study, we include an extensive set of X-ray
imaging of human lungs (CXR) with COVID-19 pneumonia, influenza virus pneumonia, and normal biomarkers to achieve an extensible and accurate CNN model.
In the experimentation phase of the research, we evaluated a variety of convolutional network architectures, selecting a sequential convolutional network with
two traditional convolutional layers and two pooling layers with maximum function. In its classification performance, the best performing model demonstrated a
validation accuracy of 93\% and an F1 score of 0.95. We chose the Azure Machine
Learning service to perform network experimentation and solution deployment.
The auto-scaling compute clusters offered a significant time reduction in network
training. We would like to see scientists across fields of artificial intelligence and
human biology collaborating and expanding on the proposed solution to provide
rapid and comprehensive diagnostics, effectively mitigating the spread of the virus.    }


  \maketitle
  
%

\section{Introduction}

   In December 2019, an epidemic caused by severe acute respiratory
syndrome coronavirus 2 (SARS-CoV-2) broke out in Wuhan, China. Coronavirus
disease, i.e. COVID-19, results from SARS-CoV-2 infection, has caused human to-human transmission (HHT) and death worldwide.

As of December 2021, global statistics, reported to the World Health
Organization, demonstrate more than 263 million confirmed cases of COVID-19,
including more than 5.23 million deaths. The new coronavirus causes severe
inflammation in human lungs, damaging the cells and tissue that line the air
sacs. The main pathologic manifestation of COVID-19 is pulmonary inflammation: radiographic manifestations vary and include ground-glass opacity (GGO), consolidation, or GGO mixed with consolidation.

Influenza is a highly contagious disease that occurs worldwide. Influenza viruses (mostly type A (H1N1), occasionally type B) cause influenza virus
pneumonia, resulting in seasonal epidemics of community-acquired pneumonia.
The main radiographic manifestations of influenza virus pneumonia are GGO and
consolidation with air bronchogram, interlobular septal thickening, centrilobular nodules, and reticular opacities (“CT Manifestations of Coronavirus Disease
(COVID-19) Pneumonia and Influenza Virus Pneumonia: A Comparative Study”, 2021).

Radiologists in China and in the United States distinguished coronavirus disease 2019 from viral pneumonia at chest radiographic pattern with moderate to high accuracy. Compared with non-COVID-19 pneumonia, COVID-19 pneumonia was more likely to have a peripheral distribution (80\% vs 57\%, P < .001), ground-glass opacity (91\% vs 68\%, P < .001), fine reticular opacity (56\% vs 22\%, P < .001), and vascular thickening (59\% vs 22\%, P < .001), but it was less likely to have a central and peripheral distribution (14\% vs 35\%, P < .001), pleural effusion (4\% vs 39\%, P < .001), or lymphadenopathy (3\% vs 10\%, P = .002)
(“Performance of Radiologists in Differentiating COVID-19 from Non-COVID-19 Viral Pneumonia at Chest CT”, 2020).

The recent study on the comparison of the radiographic manifestations of COVID-19 Pneumonia and Influenza Virus Pneumonia conducted by Lin (2021) demonstrated that the most lesions in patients with COVID-19 pneumonia were located in the peripheral zone and close to the pleura, whereas influenza
virus pneumonia was more prone to show mucoid impaction and pleural effusion. The studies conducted by Lin (2021) and Bai (2020) are aligned in their findings. However, differentiating between COVID-19 pneumonia and influenza virus pneumonia in clinical practice still remains difficult.

Therefore, we would like to develop an artificial neural network - an algorithmic approach that complements a radiological diagnosis. In our experimentation, we consider convolutional neural network topology and architectures to
determine a highly performant model classifying among normal lung biomarkers,
COVID-19 and influenza virus pneumonia images.

The chest radiograph appearance of COVID-19 pneumonia is thought
to be nonspecific, having presented a challenge to identify an optimal architecture
of a convolutional neural network that would classify with a high sensitivity among
the pulmonary inflammation features of COVID-19 and non-COVID-19 types of
pneumonia. Rahman (2021) states that COVID-19 radiography images observe
unavailability and quality issues impacting the diagnosis process and affecting the
accuracy of the deep learning detection models. A significant scarcity of COVID-19 CXR introduces an imbalance in data motivating the use of over-sampling
techniques.

In the research, we employ an extensive set of publicly available X-ray
imaging of human lungs with COVID-19 pneumonia, influenza virus pneumonia,
and healthy biomarkers. We gather data from Kaggle datasets, created by an
online community of data scientists and machine learning practitioners, and from
Mendeley Data, a secure cloud-based repository.

We chose to perform model experimentation on the Azure Machine
Learning platform since it offers a wide range of productive experiences to build,
train, and deploy machine learning models, as well as to foster team collaboration. Leveraging the auto-scaling compute feature of the Azure Machine Learning
platform allows us to manage compute resources for better training distribution,
rapid testing, and validation, as well as model deployment. As a part of the experiment design, we compared the model processing time using available compute
resources, i.e. evaluate model training time on CPU and GPU clusters.
\section{Literature Review}
In our research, we investigate recently conducted medical research and
studies on the differences in computed tomography manifestations of coronavirus
disease (COVID-19) pneumonia and those of influenza virus pneumonia. Primarily, we leverage the findings from the research “Performance of Radiologists in Differentiating COVID-19 from Non-COVID-19 Viral Pneumonia at Chest
CT” (Bai et al., 2020) and from the comparative study “CT Manifestations of
Coronavirus Disease (COVID-19) Pneumonia and Influenza Virus Pneumonia: A
Comparative Study” (Lin et al., 2021). The discoveries provide points of reference
for distinguishing SARS-CoV-2 infection from influenza virus infection based on
the CT morphologic features and quantitative parameters of COVID-19 pneumonia
and influenza virus pneumonia. However, it is stated that differentiating between
COVID-19 pneumonia and influenza virus pneumonia in clinical practice still
presents a challenge. Our study leverages two-dimensional radiography images,
we refer to the findings detected via computer tomography as a reference, and
build a knowledge base around the studied conditions.

Another study titled “Differential Diagnosis of COVID-19 Pneumonia
From Influenza A (H1N1) Pneumonia Using a Model-Based on Clinicalradiologic
Features” demonstrates analysis and a method that compares the clinicoradiologic
data of the patients with COVID-19 and H1N1 types of pneumonia. The researchers optimized the clinicoradiologic features by the least absolute shrinkage
and selection operator (LASSO) logistic regression analysis to generate a model
for differential diagnosis. They used receiver operating characteristic (ROC) curve
plots to assess the performance of the model in the primary and validation cohorts.
Their findings suggest that peripheral distribution patterns, older age, low-grade
fever, and slightly elevated aspartate aminotransferase (AST) were associated with
COVID-19 pneumonia, whereas, a peribronchovascular distribution pattern, centrilobular nodule or tree-in-bud sign, consolidation, bronchial wall thickening or
bronchiectasis, younger age, hyperpyrexia, and a higher level of AST were associated with H1N1 pneumonia. At this current stage, our research scope considers
features manifested and available through X-Rays, and is limited to considering
demographic and physiological symptoms of the COVID-19 and H1N1 pneumonia, however, the solution might be extended by enriching a feature set upon its
availability.

As for the newly proposed deep learning solutions in the researched
domain, we reference the paper “Deep Learning-Driven Automated Detection of
COVID-19 from Radiography Images: a Comparative Analysis” (Rahman et al.,
2021) since it covers challenges due to the unavailability and quality issues related
to COVID-19 radiography images impacting the diagnostic process and affecting
the accuracy of the detection model. The challenge of the unavailability speaks
to having a sufficient number of X-ray images of pneumonia-affected and normal
lungs, and a significant scarcity of COVID-19 radiography images introducing an
imbalance in data. The researchers invoked techniques of Synthetic Minority Over
Sampling (SMOTE), borderline SMOTE, and safe level SMOTE. Among deep
learning-based diagnosis approaches, the researchers discuss transfer learning,
ensemble learning, domain adaptation, cascaded networks, along with some other
approaches.

The authors are concerned about the limitations of the existing deep
convolutional neural networks like ResNet, DenseNet, and VGGNet due to having
a deep structure with excessively large parameter sets and lengthy training time.
Whereas in Transfer Learning (TL), knowledge acquired from the training on
one dataset is reused in another task with a related dataset, yielding improved
performance and faster convergence.

Chest X-ray image of a COVID-19 patient has a different distribution
but similar characteristics as that of pneumonia, allowing a promising usage of the
domain adaptation technique, i.e. using feature adversarial adaptation.

The paper speaks to the significant contributions of ensemble learning towards achieving an accurate result for COVID-19 detection as well. For
instance, Goodwin combined 12 models (Resnet-18,50,101,152, WideResnet-50,101, ResNeXt-50,101, MobileNet-v1, Densenet-121,169,201) demonstrating
better results (Goodwin et al., 2020). Similarly in the study “Pneumonia detection
in chest X-ray images using an ensemble of deep learning models” (Kundu et al.,
2021), the researchers employed deep transfer learning to handle the scarcity of
available data and designed an ensemble of three convolutional neural network
models: GoogLeNet, ResNet-18, and DenseNet-121. A weighted average ensemble technique was adopted, wherein the weights assigned to the base learners were
determined using a novel approach.

The results, discovered in the study “Deep Learning-Driven Automated
Detection of COVID-19 from Radiography Images: a Comparative Analysis”
(Rahman et al., 2021), show that the DenseNet201 model with Quadratic SVM
classifier performs the best (accuracy: 98.16\%, sensitivity: 98.93\%, specificity:
98.77\%) and maintains high accuracy in other similar architectures as well.
The recent findings on the similar detection classification problem are
promising, inspiring, and useful to our experimentation.

\section{Data Collection}
In the research, we employ publicly available X-ray imaging of human
lungs with COVID-19 pneumonia, influenza virus pneumonia, and health biomarkers. Data was collected from Kaggle datasets, created by an online community of
data scientists and machine learning practitioners, and Mendeley Data, a secure
cloud-based repository. The input data are represented by three classes:

\begin{enumerate}
\item X-ray images of human lungs with COVID-19 pneumonia (4,152)
\item X-ray images of human lungs with influenza virus pneumonia (4,494)
\item X-ray images of healthy human lungs (10,860)
\end{enumerate}

In the pre-processing step, we manually balanced the original data, arriving at an equal number of images in each class, using the smallest dataset as the baseline (COVID-19 pneumonia at 4,152 images). The classification problem considers three classes:

\begin{enumerate}
\item 	X-Ray images with healthy human lungs were labeled as 0
\item 	Influenza virus pneumonia images were labeled as 1
\item 	COVID-19 pneumonia images - labeled as 2
\end{enumerate}

\begin{figure}[hbp]
  \centering
  \begin{subfigure}{.3\columnwidth}
    \centering
    \includegraphics[width=2cm, height=2cm]{healthy.png}
    \caption*{Fig 1. Healthy}
  \end{subfigure}%
  \hfill
  \begin{subfigure}{.3\columnwidth}
    \centering
    \includegraphics[width=2cm, height=2cm]{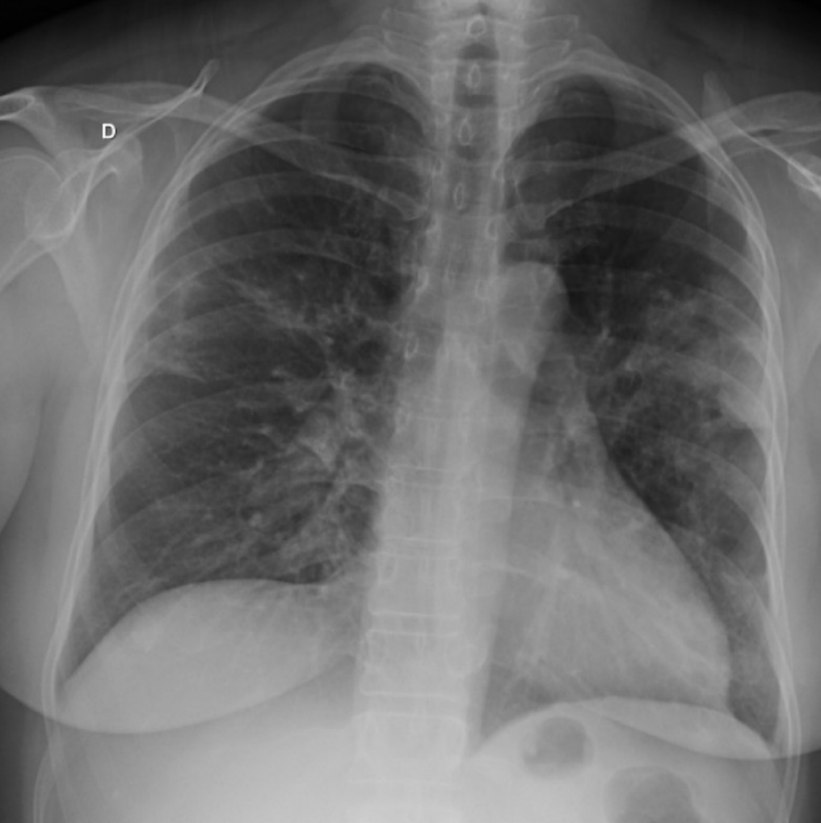}
    \caption*{Fig 2. Influenza}
  \end{subfigure}%
  \hfill
  \begin{subfigure}{.3\columnwidth}
    \centering
    \includegraphics[width=2cm, height=2cm]{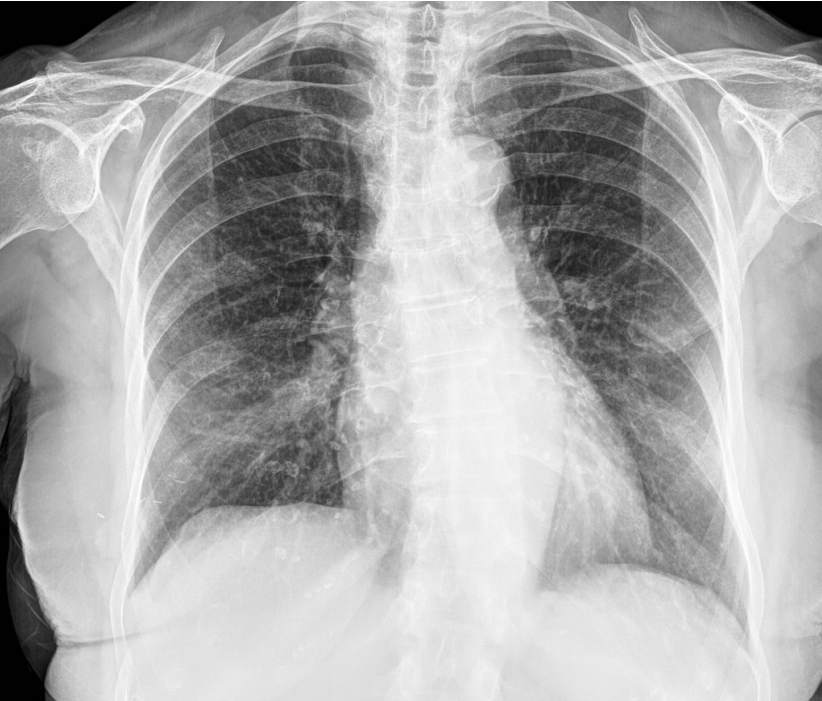}
    \caption*{Fig 3. COVID-19}
  \end{subfigure}

\end{figure}

As for image scaling, we addressed it by resizing each image to
300x300, solving the inconsistency in the image sizing. To normalize the image vectors, we divided the values within by 255. We split the data into train and test sets with 20\% of the data being
retained for the test set. The train and test datasets maintained a balance in between
each of the categories in order to prevent the overfitting or underfitting in a given
category.

The final training set considered 3,374 radiographic images of healthy
human lungs, 3,327 images of lungs infected with influenza pneumonia, and
3,359 images of lungs infected with COVID-19 pneumonia whereas the final test
set consisted of 818 images of healthy human lungs, 865 images of lungs infected
with influenza pneumonia, and 833 images of lungs infected with COVID-19
pneumonia.

Leveraging an instance of Azure Data Lake, we stored the train and test
sets under a single directory for easy access during the experimentation.

\section{Methodology}

To analyze visual imagery of CXR images with COVID-19 pneumonia, influenza virus pneumonia, and normal biomarkers, we have considered a convolutional neural network topology of deep and shallow architectures.

In the initial experimentation phase, we identified a convolutional neural network with 512 units, followed by a pooling operation for two-dimensional spatial data with a size of 2x2, and a flattened layer of 64 nodes resulting in 95\% accuracy. We use the model as an indirect baseline while constructing architectures for multi-class image classification.

Convolutional networks are a specialized type of neural network that uses a mathematical operation, convolution, in place of general matrix multiplication in at least one of their layers. Such networks are characterized with advantageous little data pre-processing. Besides input layers, hidden layers that perform convolutions, and output layers, there are special layers such as a pooling
layer with either maximum or average functions, fully connected layers, and nor malization layers. After passing through a convolutional layer, the image becomes abstracted to a feature map with a tensor shape: (number of inputs) x (feature map height) x (feature map width) x (feature map channels). Three hyperparameters control the size of the output volume of the convolutional layer: the depth, stride, and padding size. The activation function is used in the final layer of a neural
network-based classifier, mapping a vector and a specific index to a real value
(Figure 4).

   \begin{figure}[hbp]
   \centering
   \includegraphics[width=8cm]{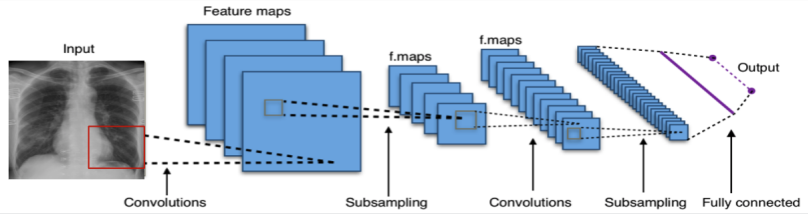}
      \caption{Figure 4. CNN Architecture}
   \end{figure}
   
In our research, we utilized the sequential convolutional network. The
approach that demonstrated the best results takes on the architecture with two
traditional convolutional layers and two pooling layers with maximum function.
The tensor of the input layer takes on a shape of 300 pixels of height, 300 pixels
of width, and 3 input channels. Convolutional layers convolve the input and
pass its result to the next layer. Max pooling layer uses the maximum value of
each local cluster of neurons in the feature map. Then, we flatten the output
of the convolutional layers to create a single long feature vector. And, the final
classification steps are performed in the fully-connected dense layers activated by
a softmax function.

Below we present the convolutional neural network architecture of the
best performing results (Figure 5):

   \begin{figure}[hbp]
   \centering
   \includegraphics[height=6cm]{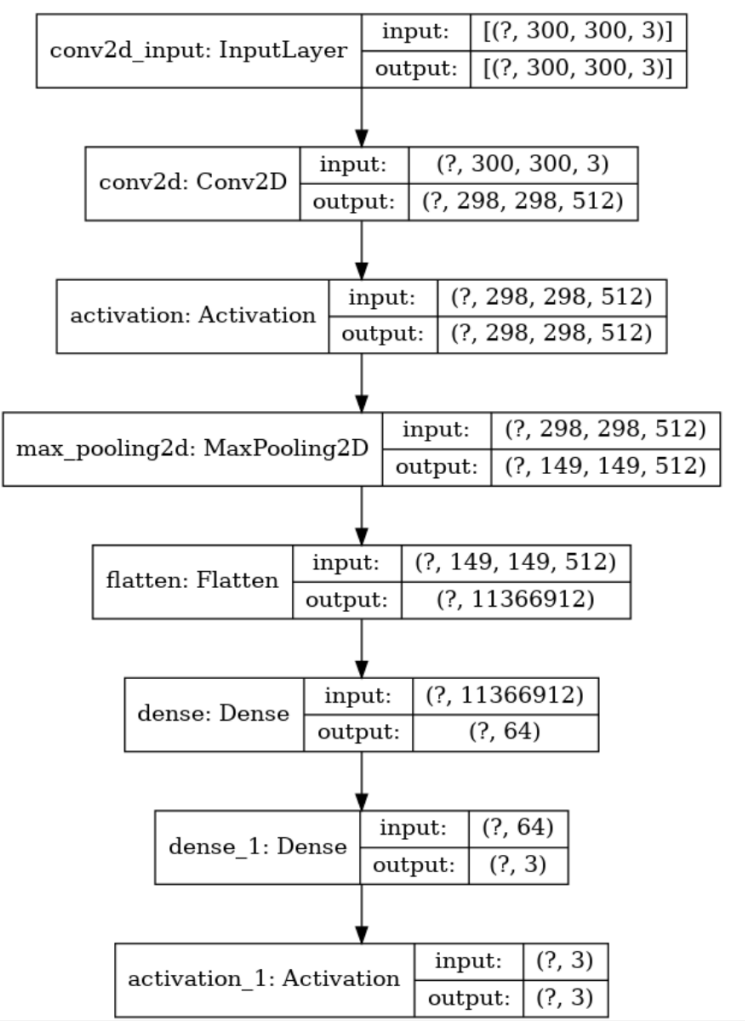}
      \caption{Figure 5. Proposed CNN}
   \end{figure}
   
\section{Results}  

The majority of the algorithms, we have evaluated, took on somewhat
shallow architectures. In the experimentation phase, among other hyperparameters, we have compared a single-layer convolutional neural network with a multilayer model, i.e. the number of convolutional layers being two. We observed a distinguishable performance difference between these two main architectures.
The analysis of the performance metrics, architecture complexity, and training time helped us to select the best model.

As the result of the model selection for the research problem, the
architecture of the best performing model takes on two convolutional layers. The
first convolutional layer is constructed with 24 filters and the second comes to 32
filters. The dense layer has 64 nodes. The processing time of the training phase
demonstrated a promising timing of 33 minutes. The model performance metrics
such as validation accuracy shows the value of 93.00\%, the validation loss results
in 0.53, and an F1 score of the most optimal model is 0.95 (Table 1).

   \begin{figure}[hbp]
   \centering
   \includegraphics[height=8cm]{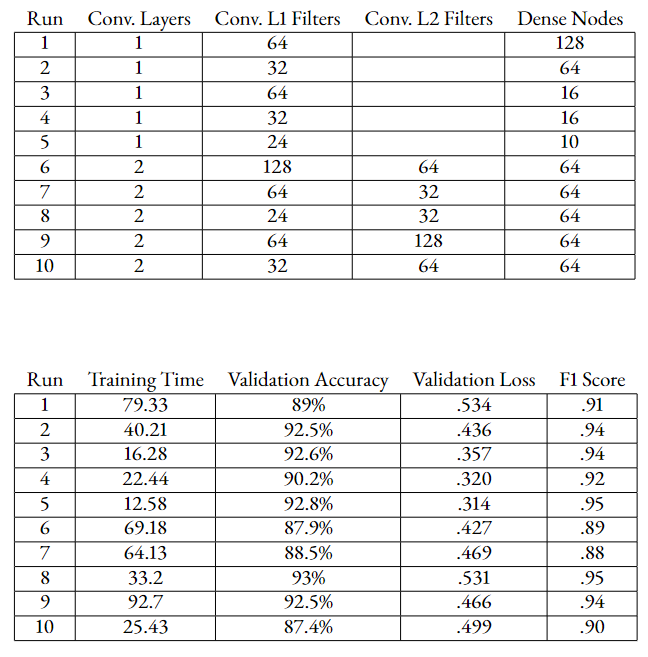}
      \caption{Table 1. Experiment details, training duration, and metrics}
   \end{figure}

To support our experimentation we leveraged Azure Machine Learning.
We performed model training using the dedicated compute clusters on the Azure
Machine Learning platform. The cluster consisted of Azure NC6 auto-scaling instances, with each instance having 6 CPU cores, 56 Gigabytes of RAM, and 1 NVIDIA Tesla K80 GPU.

We performed model testing on the balanced dataset consisting of 2,516
COVID-19 pneumonia, influenza virus pneumonia, and normal lung images. In
the confusion matrix below (Figure 6), we demonstrate the generalization abilities
of the most optimal model during the inference phase. The model classifier,
applied to the unseen data, takes on an F1 score of 0.95.
\vspace{-1\baselineskip}
   \begin{figure}[hbp]
   \centering
   \includegraphics[width=8.3cm]{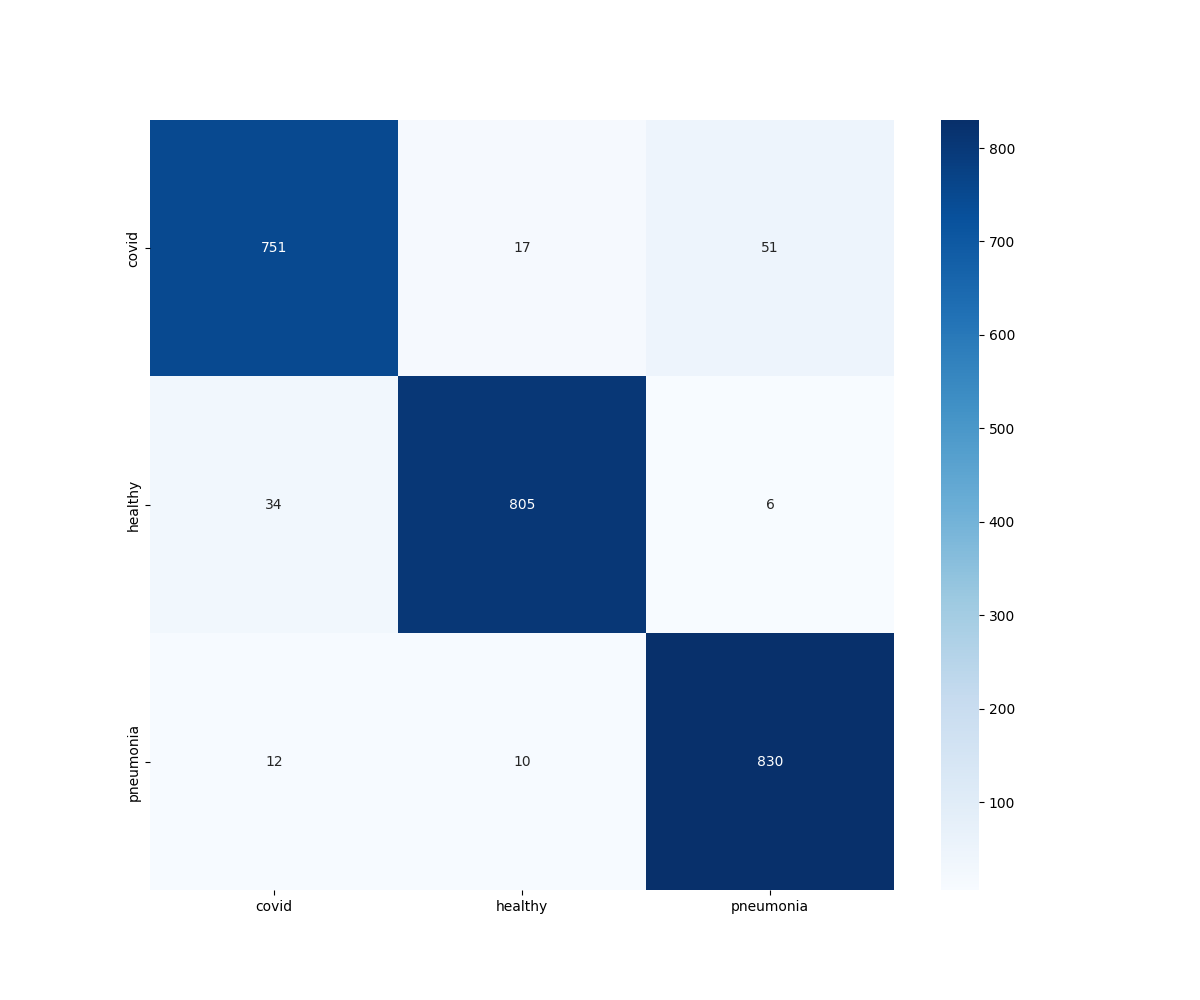}
      \caption{Figure 6. Trained CNN Confusion Matrix}
   \end{figure}
   
We leveraged the model repository of Azure Machine Learning and
deployed the model to an Azure Kubernetes Services (AKS) endpoint for real-time inference. Such an approach provides capabilities for a production-ready deployment which could be leveraged within an application development initiative, or as a standalone service to return the probability or predicted value of new chest X-ray images.

We made use of the AKS endpoint and simulated a real world application by developing a simple file upload website where radiologists and researcher could select an image for classification. The website processes the uploaded CXR image and consumes the AKS REST API endpoint as a part of the classification solution.

We performed an inference with the deployed model on 1,000 newly
collected and unseen influenza virus pneumonia images. Of these 1,000 images the
model correctly classified 979 images positive for the influenza virus pneumonia,
incorrectly classified 17 images as positive for COVID-19 pneumonia, and 4
images were detected as a class of normal lungs, resulting in an overall accuracy
of approximately 97\%. 

\section{Analysis and Interpretation}   

In summary, to accurately distinguish the visual feature representation
in COVID-19 pneumonia and influenza virus pneumonia chest X-rays, we selected
a robust deep learning architecture in computer vision, known as a convolutional neural network. The research experimentation was model-centric in nature, focusing on the hyperparameter tuning and adjusting the network architectures rather
than CXR data generation and collection methods. The research benefited from
collecting a substantial amount of relevant imagery data from public repositories. However, CXR data, gathered from disparate sources, were not consistent
observing skewed images and low resolution.

Despite high sensitivity in diagnosing, the recent study “Performance
of Radiologists in Differentiating COVID-19 from Non-COVID-19 Viral Pneumonia at Chest CT" addressed nonspecificity in the chest radiograph appearance of COVID-19 pneumonia by assessing the performance of the radiologists from the U.S. and China. Having considered 219 positive and 205 negative COVID-19 patients, the study observed four United States radiologists demonstrating high specificity percentages (93-100\%) and moderate sensitivities (73-93\%). In our research, the proposed convolutional approach shows the specificity of 97\% and
the sensitivity of 92-93\% in COVID-19 pneumonia. We also tested the proposed
method on the additional CXR images to further validate the model performance.
Our results suggest similarity to the performance in the aforementioned study by
Bai (2020).

In the research, we did not have control in the data collection methods and patient history, therefore, the model experimentation was performed only on available CXR imagery data. The additional variables such as patient symptoms, demographics, chest radiograph equipment details, evolution of patient lung states, and other factors are unknown to the scope of the study and have not been considered in the model training process. However, it is known that radiologists would leverage some or all of these factors when formulating a diagnosis. For this
reason, we view our solution as a supporting technology to the manual diagnostic
radiology, complementing opinions of radiologists.

\section{Conclusion}   

In the research, we aimed to devise a deep learning-based solution to
aid radiologists in differentiating between COVID-19 pneumonia, influenza virus
pneumonia and normal biomarkers in the CRX data. The architecture of the
highly performant model, determined in our research, yielded 93\% test accuracy
and an F1 score of 0.95, consuming only 33 minutes of compute resources during
the training phase. Our proposed solution not only demonstrates strong model
performance, but also minimizes the compute resources and time required for
future re-training efforts.

Making use of Microsoft Azure Machine Learning service, we were able to track experiment details, metrics, and charts, as well as version control of our code and trained models. The Azure Machine Learning platform offers a collaborative experience, easier tracking of model comparisons, and a streamlined model deployment process. Further, our research explores an end-to-end solution leveraging the Azure Machine Learning platform to simulate a model deployment and adaptation by radiologists. We developed a simple file upload website to
process new CXR images by submitting them to our deployed model hosted in
Azure Kubernetes Service. We used our deployed model to run inference on 1,000
unseen influenza pneumonia CXR images, resulting in 979 images being correctly
classified and an accuracy of 97\%.

Our solution is promising in complementing and supporting diagnostic radiology to differentiate between COVID-19 and influenza virus types of pneumonia, in a timely and accurate manner. We would like to see scientists, across various fields of artificial intelligence and human biology, collaborate and expand on our solution to provide rapid and comprehensive diagnostics to mitigate the spread of the virus, its mutations, variants, and strains.

\section{Directions for Future Work}  

While the current state of our research shows promising results, COVID-19 continues to mutate which may lead to shifts in pathologic manifestation and
diagnostic methods. The expanded data collection could benefit future work. In
the current research, we leveraged public repositories to gather available imagery
data which, unfortunately, observed inconsistencies and low resolution. As a next
step, we would like to collaborate with research medical facilities to gather more
relevant data on the subject (e.g. mutations, variants, strains, etc.) following with
reproducing our best performing model on the newly manifestations and feature
set. Similarly, we are excited to apply our classification network on other lung
pathologies such as chronic obstructive pulmonary disease (COPD), bronchitis,
etc. to determine its validity.

In addition to expanding available data and problem scope, we would
like to explore the impact of more complex and deep network architectures,
such as an increased number of convolutional (2 < \textit{n} <= 10) and dense layers (2 <\textit{n} <= 5). This could prove beneficial as we increase the scope of our
classifications to include variants of COVID-19, distinguish between various stages
of the disease, or identify other lung pathologies. We would also like to explore if
there would be potential benefits from dimensionality reduction techniques such as
principal component analysis (PCA) and kernel PCA, as well as additional network
regularization techniques.

From a systematic standpoint, we would like to analyze changes in
performance time by increasing processing capabilities on virtual machines with
more powerful CPUs, GPUs, and RAM. In terms of improving user accessibility
of our model, we would like to create more user-friendly mobile and desktop
platforms to complement the work of manual diagnostic radiology as well as to
allow personal utilization of the software.

%
%
\newpage
\onecolumn
\centering

\end{document}